\documentclass[]{spie}  

\usepackage{amsmath,amsfonts,amssymb}
\usepackage{graphicx}
\usepackage[colorlinks=true, allcolors=blue]{hyperref}
\usepackage{aas_macros}

\title{PERRY: A Human-in-the-Loop GUI for Precision Alignment of MUSE Data}
\author{
    Amir E. Bazkiaei$^{1,2}$, 
    Brent Miszalski$^{1,2}$, 
    Jesse van de Sande$^{3}$,
    Nuria P. F. Lorente$^{1,2}$,
    Simon~O'Toole$^{1,2}$, 
    Kate Sheng$^{1,2}$ \\
    \small $^{1}$ Australian Astronomical Optics, Faculty of Science and Engineering, Macquarie University, NSW, 2109, Australia \\
    \small $^{2}$ Astrophysics and Space Technologies Research Centre, Macquarie University, NSW, 2109, Australia \\
    \small $^{3}$ School of Physics, University of New South Wales, NSW, 2052, Australia
}

\authorinfo{Further author information: (Send correspondence to A.E.B.)\\A.E.B.: E-mail: amir.ebadati-bazkiaei@mq.edu.au}

\begin{document} 
\maketitle

\begin{abstract}
The Multi Unit Spectroscopic Explorer (MUSE) on the Very Large Telescope produces high-dimensional data-cubes where even minor astrometric misalignments between multiple exposures can introduce severe artifacts and compromise data quality. 
While automated pipelines provide a crucial baseline, subtle offsets require human intervention. 
To address this, we introduce PERRY, an interactive, Python-based Graphical User Interface (GUI) explicitly designed for the visual inspection and manual fine-tuning of MUSE image alignments. 
Playing an essential role of finding manual offset adjustments for data-cubes within the Pythonic AAO Reduction Environment (PARE), PERRY ingests automatically pre-aligned data, provides synchronized visual and quantitative diagnostics, and enables real-time coordinate translation and rotation updates. 
The resulting precise adjustments are preserved in reproducible configuration files, maximizing the scientific return and efficiency of advanced data reduction pipelines.
\end{abstract}

\section{Introduction}
The Multi Unit Spectroscopic Explorer (MUSE) \citenum{2010SPIE.7735E..08B} provides Integral-Field Spectroscopic (IFS) data-cubes that are widely used in galaxy evolution studies \citenum{2021PASA...38...31F, 2024IAUS..377...27V}. 
However, combining multiple exposures of a galaxy requires highly accurate astrometry, as misalignments can introduce artifacts, degrade spatial resolution, and weaken the visibility of morphological and structural features. 
These challenges are particularly relevant for observations of diffuse objects or regions that lack enough suitable astrometric reference sources in the field of view.  

To address this challenge, automated processing ecosystems employ specific recipes.
For example, the Pythonic AAO Reduction Environment (PARE) \citenum{2024SPIE13101E..20M} employs \texttt{MuseAlign} to automatically align MUSE images. 
\texttt{MuseAlign} utilizes \texttt{spacepylot} package \citenum{watkins_spacepylot} alignment tools while accepting a reference image as an input and automatically aligns MUSE images to the provided reference image.
However, automatic recipes, including \texttt{MuseAlign}, sometimes fail to align MUSE images properly for challenging observations. 
In such occasions, a human-in-the-loop process is necessary to visually check the images and manually align them before combining the images. 
PERRY fills this gap by functioning as an interactive human-in-the-loop step, bridging the space between fully automated pipeline processing and required visual oversight and manual alignment.

\section{Software Architecture and Pipeline Integration}
PERRY is written in Python, leveraging \textbf{PyQt6} \citenum{PyQt6} for its front-end rendering and \textbf{Matplotlib} \citenum{hunter2007matplotlib} for interactive components. 
The tool is optimized to process a batch of FITS images, which are observations of similar patch of the sky, exported by PARE's automated \texttt{MuseAlign} recipe.

\begin{figure} [ht]
   \begin{center}
   \includegraphics[width=\textwidth]{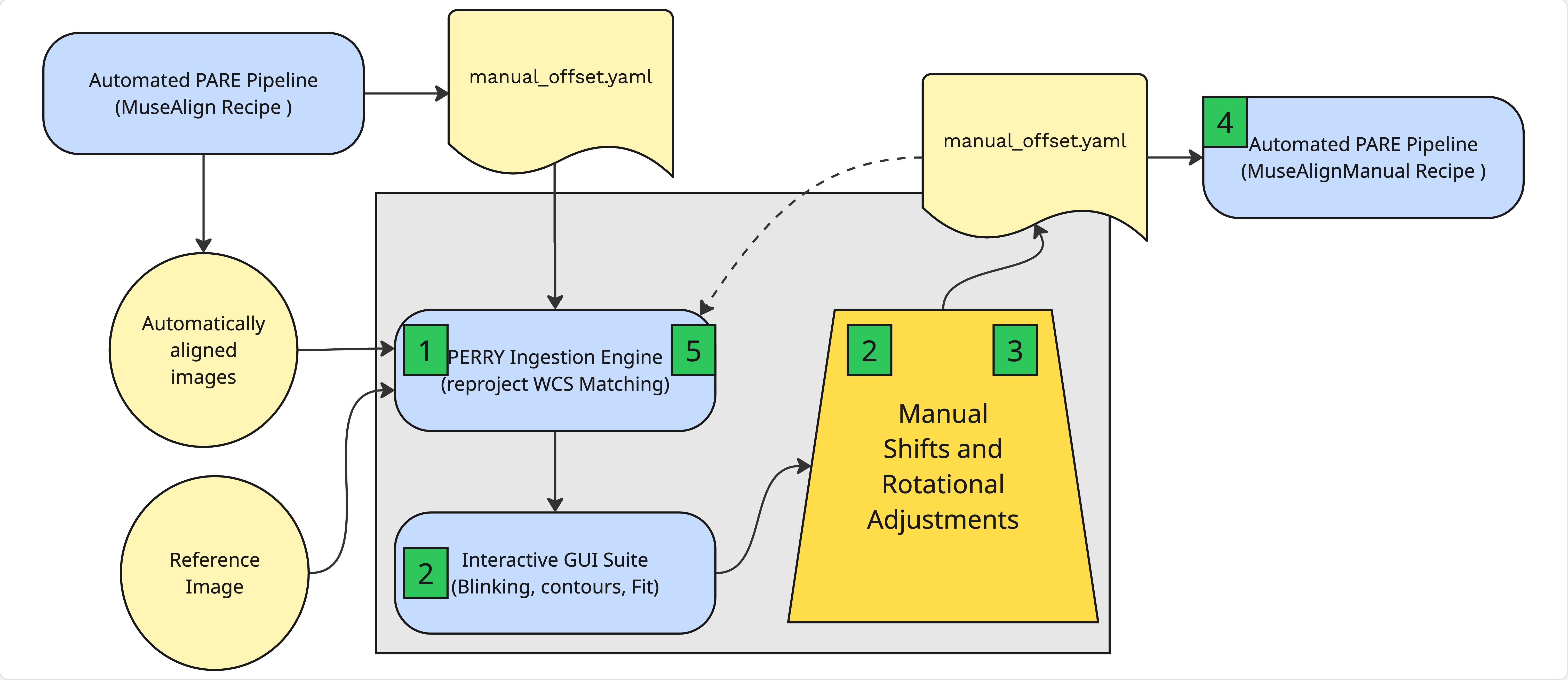}
   \end{center}
   \caption[example] 
   { \label{fig:flowchart} 
    PERRY software architecture and pipeline integration flow layout. 
    The automated \texttt{MuseAlign} recipe generates pre-aligned images and a template configuration file populated with default zero values.
    PERRY loads these alongside a reference image into the memory for interactive human inspection. 
    Green numbered labels indicate the sequence of data injection, interactive manipulation, exportation, and downstream reduction steps detailed in the main text.}
\end{figure} 

The reduction loop operates systematically as shown in Figure \ref{fig:flowchart}, where the green numbered markers denote the following processing stages:

\begin{enumerate}
    \item \textbf{Data Ingestion:} The PARE pipeline runs the \texttt{MuseAlign} recipe, which outputs automatically aligned images alongside a default template \texttt{manual\_offsets.yaml}, containing zero values as offsets. PERRY ingests these outputs along with a reference image. 
    For each MUSE image, the GUI creates a group consisting of the reference image, MUSE (aligned) image, residual image, and ratio image.
    Frame alignment is handled via standard WCS coordinate tracking utilizing the \texttt{reproject} package \citenum{2020ascl..soft11023R}.
    \item \textbf{Interactive Processing Boundary:} Active coordinate translation and rotation modifications are carried out dynamically within the GUI process memory via \texttt{scipy.ndimage} \citenum{2020NaMet..17..261V} and \texttt{skimage.transform} \citenum{2014PeerJ...2..453V} operations.
    \item \textbf{State Saving:} The manual offset transformations are stored in the \texttt{manual\_offsets.yaml} file.
    \item \textbf{Downstream Reduction:} The file is parsed by the downstream PARE recipe \texttt{MuseAlignManual}, which applies the manual offsets directly to the data-cubes prior to combining them in the next steps of PARE.
    \item \textbf{Circular Alignment Review:} Alternatively, instead of proceeding directly downstream, the system offers the flexibility to loop back into PERRY. 
    The interface is capable of parsing the contents of an existing \texttt{manual\_offsets.yaml} file; if non-zero entries from a previous PERRY session are detected, they are loaded directly into active memory. 
    This creates a circular loop, enabling users to re-load, visually inspect, or continuously refine historical manual alignments at any time.
\end{enumerate}

\section{Interface Features and Alignment Diagnostics}
The GUI coordinates three primary panels alongside navigation widgets to link visual inspection with mathematical confirmation.

\begin{figure} [ht]
   \begin{center}
   \includegraphics[width=\textwidth]{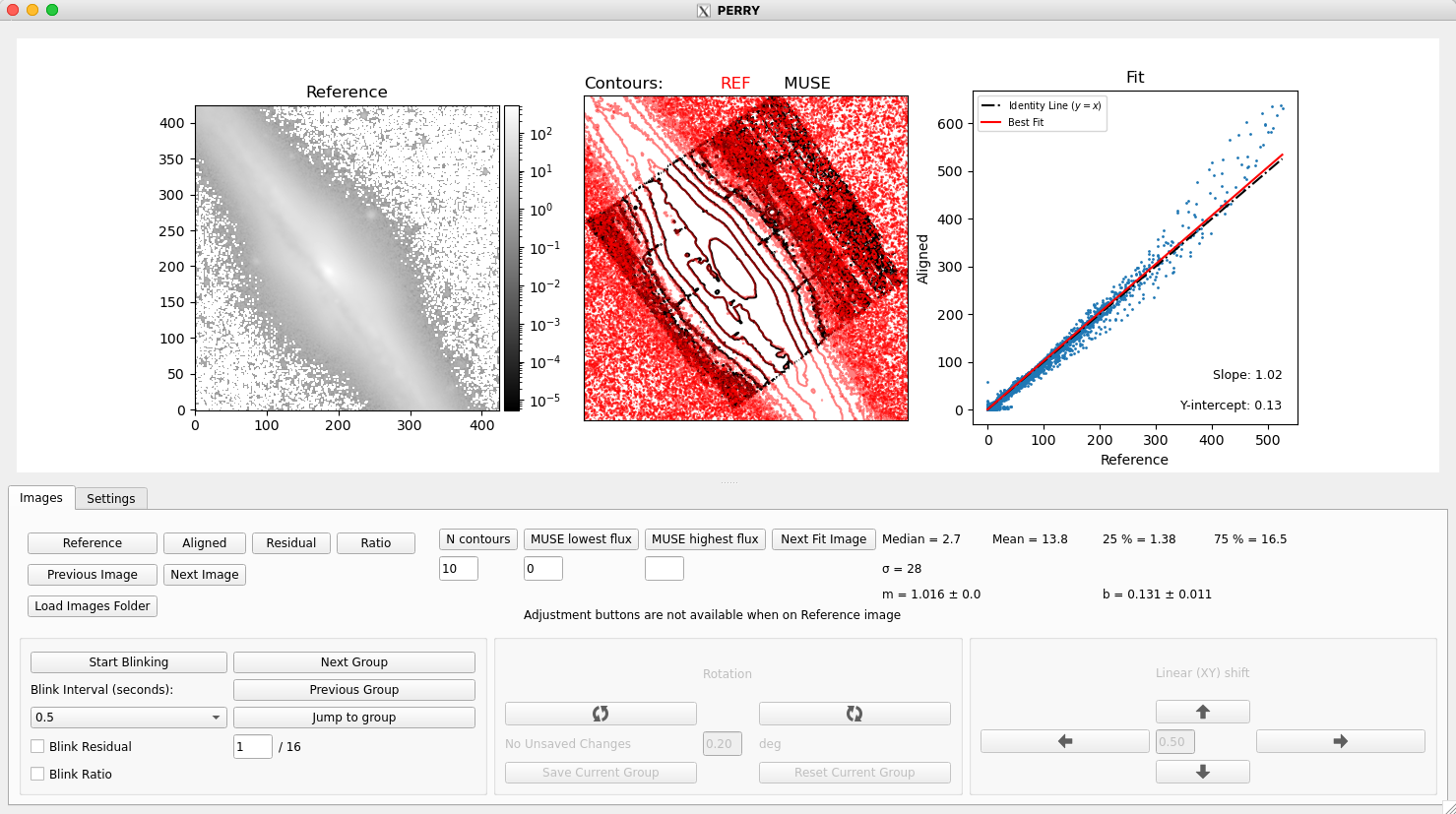}
   \end{center}
   \caption{\label{fig:gui_layout} 
    The PERRY Graphical User Interface layout during an active alignment session. 
    Displaying the primary data view the Reference frame (top, left), the dynamic contour overlay compares reference (red) and MUSE (black) structures (top, middle), and the live-updating pixel-flux linear regression plot (top, right). 
    Bottom panels: Interactive data management widgets including automated blinking configuration, group indexing (showing a batch of 16 exposures), and precision translation/rotation inputs which dynamically unlock when switching to an adjustable image state.}
\end{figure} 

The upper diagnostic panels operate synchronously as follows:
\begin{itemize}
    \item \textbf{Dynamic Visualizations and Blinking:} The main data panel (Figure \ref{fig:gui_layout}, top-left) allows users to quickly switch between the Reference, Aligned, Residual, and Ratio views. 
    An automatic blinking mode can be enabled, with the blink interval set by the user, making it easier to identify subtle spatial offsets that may not be obvious when viewing a single image. 
    The spatial display also supports smooth zooming and panning using the mouse wheel.

    \item \textbf{Contour Overlay Diagnostics:} The diagnostic panel at the top centre compares the contours of reference image (shown in red) with that of the MUSE image (shown in black). 
    As shifts and rotations are applied using the alignment controls, the contours are updated in real time, allowing the user to immediately assess the quality of the alignment.

    \item \textbf{Quantitative Fit Analysis:} The top-right panel displays a linear fit between the aligned and reference pixel fluxes, calculated using \texttt{scipy.optimize.curve\_fit} \citenum{2020NaMet..17..261V}. 
    The data are plotted against a 1:1 reference line, and the resulting slope ($m$) and y-intercept ($b$) provide a quantitative measure of the alignment quality. 
    A good alignment often will result in a close to 1:1 fit.
    Users can also restrict the flux range included in the fit by specifying lower and upper limits in the \texttt{MUSE lowest flux} and \texttt{MUSE highest flux} fields, allowing them to focus on particular structures or exclude background noise.
\end{itemize}

Beyond these primary visualization modes, the interface includes a \texttt{Settings} tab providing control over the data rendering profiles.
Users can customize the layout appearance for each of the four core state frames in the top-right panel.
Specifically, the tab enables on-the-fly modifications to the Matplotlib normalization scale (e.g., Log, Linear, Power, or SymLog), dropdown adjustments for the active colormap selection, and inputs to determine individual $v_{\min}$ and $v_{\max}$ pixel intensity thresholds.
This allows the user to optimize the visual contrast and bring out subtle high-contrast residuals dynamically during an alignment session.

\subsection{Representative User Session}
A typical manual alignment session in PERRY proceeds through the following operational workflow from start to finish:
\begin{itemize}
    \item \textbf{Initialization and Ingestion:} The user launches the interface from the terminal pointing directly to the automated reduction directory, e.g., \texttt{perry --folder path/to/MuseAlign/folder}. 
    PERRY automatically reads the automatically aligned images and template offsets file.
    \item \textbf{Baseline Assessment:} The session begins on the fixed \textit{Reference} panel layout. 
    The user compares the \textit{Reference} image to MUSE \textit{Aligned} image.
    Blinking between the \textit{Reference} image to MUSE image, comparative contours display, and the linear fit display enable the comparison.
    \item \textbf{Interactive Fine-Tuning:} The user switches the active display canvas from \textit{Reference} to \textit{Aligned} mode, unlocking the precision directional pads and rotational inputs.
    While comparing images, the user can apply fractional pixel shifts or fine angular corrections, until the structures in the \textit{Reference} image aligns to the MUSE image.
    \item \textbf{Statistical Verification:} To quantitatively validate the structural adjustment, the user observes the live regression plot (top right panel in Figure \ref{fig:gui_layout}).
    The user can apply custom thresholds to the \texttt{MUSE lowest/highest flux} input fields to only include pixels with a desired range of flux. 
    A successful manual alignment is quantitatively confirmed as the calculated slope ($m$) moves toward unity ($1.00$) and the y-intercept ($b$) drops toward zero ($0.00$) along the identity line.
    \item \textbf{Storing the offsets:} Once the alignment is completed for the current group, the user clicks \texttt{Save Current Group}.
    PERRY dynamically updates the offset file \texttt{manual\_offsets.yaml}.
    The user then utilizes the group navigation panels to cycle to the next group, repeating the validation steps for the rest of the target list.
\end{itemize}

When the manual alignment step is completed, the user runs the remaining PARE recipes.
The first PARE recipe to run after this step is \texttt{MuseAlignManual}, which applies the offets into data-cubes.

\section{Conclusion}

PERRY addresses the limitations of fully automated astrometric reduction frameworks by inserting a lightweight, human-in-the-loop validation step into the MUSE data reduction pipeline. 
By eliminating residual alignment errors prior to combining data-cubes, the interface prevents the artificial spatial smearing of astronomical structures that frequently degrades combined datasets. 
Consequently, this manual alignment step directly optimizes the sharpness of the effective point spread function (PSF) and maximizes the final scientific signal-to-noise ratio, preserving the overall integrity and structural fidelity of deep IFU observations.

The usage of PERRY extends beyond the current application to MUSE data.
Modern astronomical workflows, which exhibit a clear instrumentation trend toward next-generation IFUs on Extremely Large Telescopes, often are limited to footprints of a few tens of arcseconds (e.g. HARMONI \citenum{2024SPIE13096E..4WM, 2024arXiv241001581M}).
This spatial constraint poses severe challenges for automated astrometric recipes, particularly during high-galactic-latitude extragalactic observations where fields are naturally star-sparse.
The architectural principles implemented in PERRY are highly versatile and directly adaptable to future instruments such as HARMONI, MAVIS \citenum{10.1117/12.3018606, 2024SPIE13101E..3LF}, and BlueMUSE \citenum{2022SPIE12189E..12W} or currently available specialized spectroscopy software ecosystems like \texttt{PyKOALA} \citenum{2025arXiv250718347C}.
Ensuring robust alignment of data-cubes through interactive alignment will remain crucial for maximizing the scientific return of advanced integral field spectroscopy.

\section{Acknowledgments}

We acknowledge the Traditional Custodians of the land on which Macquarie University is situated, the Wallamattagal people of the Dharug nation, and pay our respects to Elders past and present. 
This project was funded as part of the ADACS Merit Allocation Program, under the Astronomy National Collaborative Research Infrastructure Strategy (NCRIS) Program via Astronomy Australia Ltd (AAL).

\bibliography{report} 

\begin{thebibliography}{10}

\bibitem{2010SPIE.7735E..08B}
{Bacon}, R., {Accardo}, M., {Adjali}, L., {Anwand}, H., {Bauer}, S., {Biswas},
  I., {Blaizot}, J., {Boudon}, D., {Brau-Nogue}, S., {Brinchmann}, J.,
  {Caillier}, P., {Capoani}, L., {Carollo}, C.~M., {Contini}, T., {Couderc},
  P., {Daguis{\'e}}, E., {Deiries}, S., {Delabre}, B., {Dreizler}, S.,
  {Dubois}, J., {Dupieux}, M., {Dupuy}, C., {Emsellem}, E., {Fechner}, T.,
  {Fleischmann}, A., {Fran{\c{c}}ois}, M., {Gallou}, G., {Gharsa}, T.,
  {Glindemann}, A., {Gojak}, D., {Guiderdoni}, B., {Hansali}, G., {Hahn}, T.,
  {Jarno}, A., {Kelz}, A., {Koehler}, C., {Kosmalski}, J., {Laurent}, F., {Le
  Floch}, M., {Lilly}, S.~J., {Lizon}, J.-L., {Loupias}, M., {Manescau}, A.,
  {Monstein}, C., {Nicklas}, H., {Olaya}, J.-C., {Pares}, L., {Pasquini}, L.,
  {P{\'e}contal-Rousset}, A., {Pell{\'o}}, R., {Petit}, C., {Popow}, E.,
  {Reiss}, R., {Remillieux}, A., {Renault}, E., {Roth}, M., {Rupprecht}, G.,
  {Serre}, D., {Schaye}, J., {Soucail}, G., {Steinmetz}, M., {Streicher}, O.,
  {Stuik}, R., {Valentin}, H., {Vernet}, J., {Weilbacher}, P., {Wisotzki}, L.,
  and {Yerle}, N., ``{The MUSE second-generation VLT instrument},'' in [{\em
  Ground-based and Airborne Instrumentation for Astronomy
  III}{\nolinebreak\hspace{0.1em}]},  {McLean}, I.~S., {Ramsay}, S.~K., and
  {Takami}, H., eds., {\em Society of Photo-Optical Instrumentation Engineers
  (SPIE) Conference Series} {\bf 7735},  773508 (July 2010).
\newblock DOI: \url{https://doi.org/10.1117/12.856027}.

\bibitem{2021PASA...38...31F}
{Foster}, C., {Mendel}, J.~T., {Lagos}, C.~D.~P., {Wisnioski}, E., {Yuan}, T.,
  {D'Eugenio}, F., {Barone}, T.~M., {Harborne}, K.~E., {Vaughan}, S.~P.,
  {Schulze}, F., {Remus}, R.-S., {Gupta}, A., {Collacchioni}, F., {Khim},
  D.~J., {Taylor}, P., {Bassett}, R., {Croom}, S.~M., {McDermid}, R.~M.,
  {Poci}, A., {Battisti}, A.~J., {Bland-Hawthorn}, J., {Bellstedt}, S.,
  {Colless}, M., {Davies}, L.~J.~M., {Derkenne}, C., {Driver}, S.,
  {Ferr{\'e}-Mateu}, A., {Fisher}, D.~B., {Gjergo}, E., {Johnston}, E.~J.,
  {Khalid}, A., {Kobayashi}, C., {Oh}, S., {Peng}, Y., {Robotham}, A.~S.~G.,
  {Sharda}, P., {Sweet}, S.~M., {Taylor}, E.~N., {Tran}, K.-V.~H., {Trayford},
  J.~W., {van de Sande}, J., {Yi}, S.~K., and {Zanisi}, L., ``{The MAGPI
  survey: Science goals, design, observing strategy, early results and
  theoretical framework},'' {\em \pasa}~{\bf 38},  e031 (July 2021).
\newblock DOI: \url{https://doi.org/10.1017/pasa.2021.25}.

\bibitem{2024IAUS..377...27V}
{van de Sande}, J., {Fraser-McKelvie}, A., {Fisher}, D.~B., {Martig}, M.,
  {Hayden}, M.~R., and {Geckos Survey Collaboration}, ``{GECKOS: Turning galaxy
  evolution on its side with deep observations of edge-on galaxies},'' in [{\em
  Early Disk-Galaxy Formation from JWST to the Milky
  Way}{\nolinebreak\hspace{0.1em}]},  {Tabatabaei}, F., {Barbuy}, B., and
  {Ting}, Y.-S., eds., {\em IAU Symposium} {\bf 377},  27--33 (Jan. 2024).
\newblock DOI: \url{https://doi.org/10.1017/S1743921323001138}.

\bibitem{2024SPIE13101E..20M}
{Miszalski}, B., {Bazkiaei}, A.~E., {van de Sande}, J., {O'Toole}, S.~J.,
  {Horton}, A., and {Tocknell}, J., ``{Extensible pipeline development powered
  by PyCPL and PyEsorex},'' in [{\em Software and Cyberinfrastructure for
  Astronomy VIII}{\nolinebreak\hspace{0.1em}]},  {Ibsen}, J. and {Chiozzi}, G.,
  eds., {\em Society of Photo-Optical Instrumentation Engineers (SPIE)
  Conference Series} {\bf 13101},  1310120 (July 2024).
\newblock DOI: \url{https://doi.org/10.1117/12.3013046}.

\bibitem{watkins_spacepylot}
Watkins, E.~J., ``spacepylot,'' (2026).
\newblock GitHub repository, accessed 24 June 2026,
  \url{https://github.com/ejwatkins-astro/spacepylot}.

\bibitem{PyQt6}
Limited, R.~C., ``Pyqt6: Python bindings for the qt application framework.''
  \url{https://www.riverbankcomputing.com/software/pyqt/} (2026).
\newblock Accessed 2026-06-21.

\bibitem{hunter2007matplotlib}
Hunter, J.~D., ``Matplotlib: A 2d graphics environment,'' {\em Computing in
  Science \& Engineering}~{\bf 9}(3),  90--95 (2007).

\bibitem{2020ascl..soft11023R}
{Robitaille}, T., {Deil}, C., and {Ginsburg}, A., ``{reproject: Python-based
  astronomical image reprojection}.'' Astrophysics Source Code Library, record
  ascl:2011.023 (Nov. 2020).

\bibitem{2020NaMet..17..261V}
{Virtanen}, P., {Gommers}, R., {Oliphant}, T.~E., {Haberland}, M., {Reddy}, T.,
  {Cournapeau}, D., {Burovski}, E., {Peterson}, P., {Weckesser}, W., {Bright},
  J., {van der Walt}, S.~J., {Brett}, M., {Wilson}, J., {Millman}, K.~J.,
  {Mayorov}, N., {Nelson}, A. R.~J., {Jones}, E., {Kern}, R., {Larson}, E.,
  {Carey}, C.~J., {Polat}, {\.I}., {Feng}, Y., {Moore}, E.~W., {VanderPlas},
  J., {Laxalde}, D., {Perktold}, J., {Cimrman}, R., {Henriksen}, I.,
  {Quintero}, E.~A., {Harris}, C.~R., {Archibald}, A.~M., {Ribeiro}, A.~H.,
  {Pedregosa}, F., {van Mulbregt}, P., and {SciPy 1. 0 Contributors}, ``{SciPy
  1.0: fundamental algorithms for scientific computing in Python},'' {\em
  Nature Medicine}~{\bf 17},  261--272 (Feb. 2020).
\newblock DOI: \url{https://doi.org/10.1038/s41592-019-0686-2}.

\bibitem{2014PeerJ...2..453V}
{van der Walt}, S., {Sch{\"o}nberger}, J.~L., {Nunez-Iglesias}, J., {Boulogne},
  F., {Warner}, J.~D., {Yager}, N., {Gouillart}, E., {Yu}, T., and
  {scikit-image Contributors}, ``{scikit-image: Image processing in Python},''
  {\em PeerJ}~{\bf 2},  e453 (Jan. 2014).
\newblock DOI: \url{https://doi.org/10.7717/peerj.453}.

\bibitem{2024SPIE13096E..4WM}
{Muslimov}, E.~R., {Castillo-Dom{\'\i}nguez}, E., {Kariuki}, J., {Chao-Ortiz},
  J., {Tecza}, M., {Meyer}, E., {Ozer}, Z., {Clarke}, F., and {Thatte}, N.,
  ``{HARMONI at ELT: tolerance analysis and expected as-build imaging
  performance of the infrared spectrograph},'' in [{\em Ground-based and
  Airborne Instrumentation for Astronomy X}{\nolinebreak\hspace{0.1em}]},
  {Bryant}, J.~J., {Motohara}, K., and {Vernet}, J. R.~D., eds., {\em Society
  of Photo-Optical Instrumentation Engineers (SPIE) Conference Series} {\bf
  13096},  130964W (July 2024).
\newblock DOI: \url{https://doi.org/10.1117/12.3020119}.

\bibitem{2024arXiv241001581M}
{Muslimov}, E., {Castillo-Dominguez}, E., {Kariuki}, J., {Chao-Ortiz}, J.,
  {Tecza}, M., {Meyer}, E., {Ozer}, Z., {Clarke}, F., and {Thatte}, N.,
  ``{HARMONI at ELT: tolerance analysis and expected as-build imaging
  performance of the infrared spectrograph},'' {\em arXiv e-prints} ,
  arXiv:2410.01581 (Oct. 2024).
\newblock DOI: \url{https://doi.org/10.48550/arXiv.2410.01581}.

\bibitem{10.1117/12.3018606}
Ellis, S., Zhelem, R., Horton, A., Chin, T., Fernando, N., Hewlett, M.,
  Lorente, N., Luo, S., McDermid, R., McGregor, H., Robertson, D., Schwab, C.,
  Smedley, S., Waller, L., Zheng, J., Kosmalski, J., Cresci, G., Mendel, T.,
  Bianco, A., Brodrick, D., Burgess, J., Haynes, D., Greggio, D., Content, R.,
  and Rigaut, F., ``{MAVIS: imager and spectrograph},'' in [{\em Ground-based
  and Airborne Instrumentation for Astronomy X}{\nolinebreak\hspace{0.1em}]},
  Bryant, J.~J., Motohara, K., and Vernet, J. R.~D., eds.,  {\bf 13096},
  130961B, International Society for Optics and Photonics, SPIE (2024).
\newblock \url{https://doi.org/10.1117/12.3018606}.

\bibitem{2024SPIE13101E..3LF}
{Fernando}, N., {Lorente}, N., {McDermid}, R., {Horton}, A.~J., and {Luvaul},
  L., ``{MAVIS: The preliminary data reduction pipeline},'' in [{\em Software
  and Cyberinfrastructure for Astronomy VIII}{\nolinebreak\hspace{0.1em}]},
  {Ibsen}, J. and {Chiozzi}, G., eds., {\em Society of Photo-Optical
  Instrumentation Engineers (SPIE) Conference Series} {\bf 13101},  131013L
  (July 2024).
\newblock DOI: \url{https://doi.org/10.1117/12.3019458}.

\bibitem{2022SPIE12189E..12W}
{Weilbacher}, P.~M., {Martens}, S., {Wendt}, M., {Roth}, M.~M., {Dreizler}, S.,
  {Kelz}, A., {Bacon}, R., and {Richard}, J., ``{The BlueMUSE data reduction
  pipeline: lessons learned from MUSE and first design choices},'' in [{\em
  Software and Cyberinfrastructure for Astronomy
  VII}{\nolinebreak\hspace{0.1em}]},  {\em Society of Photo-Optical
  Instrumentation Engineers (SPIE) Conference Series} {\bf 12189},  1218912
  (Aug. 2022).
\newblock DOI: \url{https://doi.org/10.1117/12.2630024}.

\bibitem{2025arXiv250718347C}
{Corcho-Caballero}, P., {Ascasibar}, Y., {L{\'o}pez-S{\'a}nchez}, {\'A}.~R.,
  {Gonz{\'a}lez-Bolivar}, M., {Lorente}, N. P.~F., {Tocknell}, J.,
  {Jim{\'e}nez-Ibarra}, F., {Jayasuriya Daluwathumullagamage}, P.,
  {Quattropani}, G., {Owers}, M., and {Verdoes-Kleijn}, G.~A., ``{The PyKOALA
  python library: a multi-instrument package for IFS data reduction},'' {\em
  arXiv e-prints} ,  arXiv:2507.18347 (July 2025).
\newblock DOI: \url{https://doi.org/10.48550/arXiv.2507.18347}.

\end{thebibliography}
\bibliographystyle{spiebib} 

\end{document}